\documentclass{aa}
\usepackage{graphicx}
\usepackage{txfonts}

\begin{document} 

\title{Thermal desorption of circumstellar and cometary ice analogs}

\author{R. Mart\'in-Dom\'enech \inst{\ref{inst1}} \and G.M. Mu\~noz Caro \inst{\ref{inst1}} \and J. Bueno \inst{\ref{inst2}} 
\and F. Goesmann \inst{\ref{inst3}}}

\institute{Centro de Astrobiolog\'ia (INTA-CSIC), Ctra. de Ajalvir, km 4, Torrej\'on de Ardoz, 28850 Madrid, Spain
\email{rmartin@cab.inta-csic.es}\label{inst1}
\and
Netherlands Institute for Space Research, Sorbonnelaan 2, 3584 CA Utrecht, Netherlands \label{inst2}
\and
Max Planck Institute for Solar System Research, Justus von Liebig Weg 3, 370077 G\"ottingen, Germany \label{inst3}}

\date{}

\abstract
{Thermal annealing of interstellar ices takes place in several stages of star formation. 
Knowledge of this process comes from a combination of astronomical observations and laboratory simulations under astrophysically relevant conditions.}
{For the first time we present the results of temperature programmed desorption (TPD) 
experiments with pre-cometary ice analogs composed of up to five molecular components: H$_{2}$O, CO, CO$_{2}$, CH$_{3}$OH, and NH$_{3}$.}
{The experiments were performed with an ultra-high vacuum chamber.  
A gas line with a novel design allows the controlled preparation of mixtures with up to five molecular components. 
Volatiles desorbing to the gas phase were monitored using a quadrupole mass spectrometer, while changes in the ice structure and composition were studied 
by means of infrared spectroscopy.}
{The TPD curves of water ice containing CO, CO$_{2}$, CH$_{3}$OH, and NH$_{3}$ present desorption peaks at temperatures near those observed in pure ice experiments, 
volcano desorption peaks after water ice crystallization, and co-desorption peaks with water. 
Desorption peaks of CH$_{3}$OH and NH$_{3}$ at temperatures similar to the pure ices 
takes place when their abundance relative to water is above $\sim$ 3 \% in the ice matrix. 
We found that CO, CO$_{2}$, and NH$_{3}$ also present co-desorption peaks with CH$_{3}$OH, which cannot be reproduced in experiments with 
binary water-rich ice mixtures. These are extensively used in the study of thermal desorption of interstellar ices. 
}
{These results reproduce the heating of circumstellar ices in hot cores  
and can be also applied to the late thermal evolution of comets.   
In particular, TPD curves represent a benchmark for the analysis of the measurements that mass spectrometers on board the ESA-Rosetta cometary mission will perform 
on the coma of comet 67P/Churyumov-Gerasimenko, which will be active before the arrival of Rosetta according to our predictions. 
}
\keywords{ISM: ices - ISM: hot cores - solar system - methods: laboratory - infrared spectroscopy - mass spectrometry}

\maketitle

\section{Introduction}
\label{intro}

Dense molecular clouds have typical densities of 10$^{4}$-10$^{6}$ particles cm$^{-3}$ and temperatures down to 10 K in their interiors. 
These low temperatures are reached thanks to the screening from the interstellar ultraviolet (UV) radiation field provided by 
the gas, dust particles and, presumably, polycyclic aromatic hydrocarbons (PAHs) at the edge of the cloud that absorb most of the radiation. 
Under these conditions,  molecules are able to condense onto the surface of dust grains, forming ice mantles. 
The size of interstellar dust grains is usually given by dust models as a power-law distribution, covered by a $\sim$ 0.01 
$\mu$m ice mantle in dense cloud interiors (Zubko et al. 2004).

When a dense core collapses to form a star, 
grain agglomeration can take place during the cold core phase, leading to cluster particles composed of sub-micron grains. 
Later on, during the warm-up phase, 
the protostar heats its environment,  
leading to thermal processing of ice mantles.  
This thermal annealing takes place in two different ways.
On the one hand, grains at a certain distance are heated 
(approximately 1 K/century; Viti \& Williams 1999) 
as the temperature of the protostar increases. 
On the other hand, grains 
are also able to undergo periodic radial excursions to distances less than 0.1 AU from the central protostar on short timescales 
(from a few to hundreds of hours), 
according to the fluctuating X-wind model for the formation of Ca-Al-rich inclusions (CAIs) and chondrules (Shu et al. 1996, 1997, 2001).

The main effect of thermal annealing is the sublimation of ice mantles explaining 
the excess gas-phase abundances of some species 
in the inner regions of protostellar envelopes (e.g., Pauls et al. 1983; Herbst \& van Dischoeck 2009 and ref. therein).
Models predict that thermal desorption of ice governs the gas-phase chemistry during star formation (e.g., Viti et al. 2004). 

The remaining ice mantles in protoplanetary disks are incorporated in comets and other minor bodies. 
Cometary ice composition is thought to be 
similar to that of the interstellar ice mantles (Mumma \& Charnley 2011 and ref. therein), although it can vary between comets. 
Molecules present on ices, both originally or created afterwards 
by means of 
energetic processing, are thus transported by these bodies, which may be 
responsible for the delivery of organic material to habitable planets, and, ultimately, the origin of life.
 
Cometary ices are also exposed to thermal processing, which is the main responsible for the activity observed in comets.
When a comet passes within r $\sim$ 3 AU of the Sun during its orbit, 
icy volatiles form an expanding atmosphere of gas and dust
called the coma (Wyckof 1982). 

Most of our current knowledge on interstellar, circumstellar, and cometary ices comes from a combination of infrared observations and laboratory experiments 
that simulate their energetic processing under astrophysically relevant conditions. 
Heating of ice mixtures leads to structural changes in the ice, involving diffusion of molecules,  
phase changes, and ice segregation. 
Segregation is expected to occur whenever ice diffusion of molecules is possible, and it is energetically favorable for molecules of the same kind 
to group together.  
Ice diffusion barriers are proportional to the binding energy of the species (i.e., their volatility). 
Thermal annealing of ice samples also leads to sequential desorption of ice molecules that are able to diffuse through the structure of the ice toward the surface,    
which starts with the most volatile species (\"Oberg et al. 2009).  

The experiments presented here contribute to 
a better understanding on the thermal processing of circumstellar and cometary ices.
The majority of published results of temperature programmed desorption (TPD) experiments of astrophysical ice analogs 
deal with pure ices or binary mixtures with H$_{2}$O as the primary component (e.g., Collings et al. 2004). 
Some ternary mixtures have been treated as well (e.g., Fayolle et al. 2011).\\
This work represents the first attempt to study the thermal processing of a more realistic astrophysical ice with up to 
five molecular components (H$_{2}$O, CO, CO$_{2}$, CH$_{3}$OH, and NH$_{3}$) using both mass spectrometry  
and infrared spectroscopy. 

Our results can also be used to predict and analyze  
the data collected by mass spectrometers 
on board cometary missions like Rosetta. 
The International Rosetta Mission was approved in 1993 and has been developed by the European Space Agency (ESA).  
Since its approval, several works have been published studying the target of this cometary mission using both ground and space-based observations 
with special interest on the evolution of the activity during its orbit  
(e.g., Kelley et al. 2009; de Almeida et al. 2009; Snodgrass et al. 2013).
Rosetta was launched in March 2004 and will arrive at comet 
67P/Churyumov-Gerasimenko (a short-period comet with P = 6.45 yrs) 
in July 2014 at $\sim$ 3.8 AU from the Sun. 
The Rosetta orbiter will remain in close proximity to the comet nucleus, and a small lander will be released onto its surface. 
When the mission draws to a close in December 2015, the comet will be on its way out of the inner Solar System.
Previous cometary missions have provided in situ measurements of cometary gas composition (e.g., for comet Halley; Eberhardt 1999). 
Ground-based observations have been also used to detect coma components (Mumma \& Charnley 2011 and ref. therein). 
However, no direct measurements of cometary ice composition are avaliable so far.
The NASA-Deep Impact cometary mission studied the nucleus of comet 9P/Temple 1 but from a more geological point of view (e.g., Thomas et al. 2007). 
Rosetta instruments will provide for the first time simultaneous measurements of both the nucleus and the coma of a comet. 
Our experiments, using realistic astrophysical ice analogs, can be used to understand the relation between the nucleus and the coma compositions 
and also the cometary ice conditions. 
However, as we discuss, caution is needed to extrapolate these experimental results to a real cometary scenario.

The layout of this paper is as follows. In Sect. 2, we describe the experimental protocol. Section 3 presents the experimental results. Their astrophysical 
implications are discussed in Sect. 4. Section 5 summarizes the main conclusions. 

\section{Experimental}

The results presented here have been obtained using the novel InterStellar Astrochemistry Chamber (ISAC) 
at the Centro de Astrobiolog\'ia (Mu\~noz Caro et al. 2010). 
The ISAC set-up is an ultra-high vacuum (UHV) chamber with a pressure 
about 4 x 10$^{-11}$ mbar, which corresponds to a density of 10$^{6}$ cm$^{-3}$ (Mu\~noz Caro et al. 2010), similar to that found in dense cloud interiors. 
Ice samples made by deposition of a gas mixture onto a KBr window 
at 8 K (achieved by means of a closed-cycle helium cyostat) 
were warmed up until a complete sublimation was attained. 
A silicon diode temperature sensor and a LakeShore Model 331 temperature controller are used, reaching a sensitivity of about 0.1 K.
Complex gas mixtures were prepared in the gas line system, using 
a Pfeiffer Prisma quadrupole mass spectrometer (QMS) of mass spectral range from 1 to 100 amu with a Faraday detector, 
and electrovalves were used to control the flow of the components, thus allowing co-deposition of gas mixtures with the desired 
composition. 
A second deposition tube was used to introduce corrosive gases, such as NH$_{3}$. 

The chemical components used in the experiments were H$_{2}$O (liquid), triply distilled, CH$_{3}$OH (liquid, Panreac 99.9\%), CO (gas, Praxair 99.998\%), 
CO$_{2}$ (gas, Praxair 99.996\%), and NH$_{3}$ (gas, Praxair 99.999\%),  
which was deposited through the second deposition tube. 

The evolution of the solid sample was monitored by in situ Fourier transform infrared (FTIR) transmittance spectroscopy, 
using a Bruker Vertex 70 spectrometer equipped with a deuterated triglycine sulfate detector (DTGS). 
The IR spectra were collected after ice deposition at 8 K, or every five minutes during warm-up, with a spectral resolution of 2 cm$^{-1}$.
Column densities of each species in the ice were calculated from the IR spectra using the formula

\begin{equation}
N=\frac{1}{A}\int_{band}{\tau_{\nu} \ d\nu},
\end{equation}

where $N$ is the column density in molecules cm$^{-2}$, $\tau_{\nu}$ the optical depth of the absorption band, and $A$ the band strength in cm molecule$^{-1}$, 
as derived from laboratory experiments (Table \ref{fuerzas}). 
Band strengths were measured for pure ices made of one molecular component. The same values are usually adopted in ice mixtures, which introduce an 
uncertainty of about 20-30\% (d'Hendecourt \& Allamandola 1986). 

\begin{table}
\begin{tabular}{|c|c|c|c|c|}
\hline
Molecule&Vibrational&Frequency&Band strength\\
&mode&(cm$^{-1}$)&(cm molec$^{-1}$)\\
\hline
H$_{2}$O&stretching O-H&3280&2.0$\times 10^{-16}$ $^{a}$\\
\hline
CO&stretching C=O&2138&1.1$\times 10^{-17}$ $^{b}$\\
\hline
CO$_{2}$&stretching C=O&2344&7.6$\times 10^{-17}$ $^{c}$\\
\hline
CH$_{3}$OH&stretching C-O&1025&1.8$\times 10^{-17}$ $^{d}$\\
\hline
NH$_{3}$&umbrella mode&1070&1.7$\times 10^{-17}$ $^{e}$\\
\hline
\end{tabular}
\caption{IR feature used to calculate the column density of each component (frequencies and band strengths for pure ices at 8 K).}
\begin{footnotesize}
{a From Hagen 1981\\
b From Jiang et al. 1975\\
c From Yamada \& Person 1964\\
d From d'Hendecourt \& Allamandola 1986\\
e From Sandford \& Allamandola 1993}
\end{footnotesize}
\label{fuerzas}
\end{table}

The desorbing 
molecules were detected by a second Pfeiffer Prisma QMS of mass spectral range from 1 to 200 amu with a Channeltron detector, 
which is situated $\sim$ 17 cm apart from the sample.  
Gas-phase molecules were ionized by low energy ($\sim$ 70 eV) electron bombardment.
Every species was monitored 
through its main mass fragment (except for NH$_{3}$), namely: 
m/z=18 (H$_{2}$O), 
m/z=28 (CO, with a small contribution of CO$_{2}$ fragmentation into CO$^{+}$ of, approximately, 10\% of the m/z=44 signal), 
m/z=31 (CH$_{3}$OH, because the ion molecule with m/z=32 
coincides with O$_{2}$), and
m/z=44 (CO$_{2}$). 
Ammonia, NH$_{3}$, was monitored through the mass fragment m/z=15 to avoid confusion with the mass fragment m/z=17 of H$_{2}$O.  
The m/z=17 signal was, nonetheless, used to confirm that the m/z=15 signal was not produced by contaminants present in small amounts.

An overview of all the experiments is shown in Table \ref{exp}. 
All ice analogs have column densities above one hundred monolayers (1 ML is commonly defined as 10$^{15}$ molecules/cm$^{2}$), which are well above the canonical thickness 
of a thin interstellar ice mantle ($\sim$ 40 ML; Bisschop et al. 2006). 
Therefore, these experiments  more faithfully reproduce the thermal annealing of thicker ice mantles 
formed in cold circumstellar regions 
by grain agglomeration 
(and even cometary ices, see Sect. \ref{imp}).  
Pre-cometary ice analog composition in experiments E10 and E11 is similar to that found in most interstellar and pre-cometary icy environments 
(Mumma \& Charnley 2011 and ref. therein). 
We also performed additional experiments (E6-E9) involving ices with a different composition 
to study in depth some particular effects. 
Experiments with pure ices of each component were performed as references to study mixture effects (E1-E5).

\begin{table}
\begin{tabular}{c|c|c|c|c|c|c}
Exp&H$_{2}$O&CO&CO$_{2}$&CH$_{3}$OH&NH$_{3}$&N/10$^{15}$\\
&(\%)&(\%)&(\%)&(\%)&(\%)&mol/cm$^{2}$\\
\hline
\hline
E1 $^{a}$&100.0&0.0&0.0&0.0&0.0&1554.61\\
&100.0&0.0&0.0&0.0&0.0&\\
\hline
E2 $^{a}$&0&100.0&0.0&0.0&0.0&104.97\\
&0.0&100.0&0.0&0.0&0.0&\\
\hline
E3 $^{a}$&0.0&0.0&100.0&0.0&0.0&134.88\\
&0.0&0.0&100.0&0.0&0.0&\\
\hline
E4 $^{a}$&0.0&0.0&0.0&100.0&0.0&190.81\\
&0.0&0.0&0.0&100.0&0.0&\\
\hline
E5 $^{a}$&0.0&0.0&0.0&0.0&100.0&154.60\\
&0.0&0.0&0.0&0.0&100.0&\\
\hline
\hline
E6 $^{b}$&100.0&7.8&6.8&1.6&0.0&308.69\\
&86.1&6.7&5.8&1.3&0.0&\\
\hline
E7 $^{b}$&100.0&6.6&27.9&3.1&3.4&1395.54\\
&70.9&4.7&19.8&2.2&2.4&\\
\hline
E8 $^{b}$&100.0&0.0&0.0&9.3&0.0&109.52\\
&91.5&0.0&0.0&8.5&0.0&\\
\hline
E9 $^{b}$&100.0&0.0&0.0&68.6&0.0&385.58\\
&59.3&0.0&0.0&40.7&0.0&\\
\hline
\hline
E10 $^{b}$&100.0&8.7&20.2&5.7&5.7&749.50\\
&71.2&6.2&14.4&4.1&4.1&\\
\hline
E11 $^{a}$&100.0&15.0&10.3&7.2&8.0&912.43\\
&71.2&10.7&7.3&5.7&5.1&\\
\hline
\end{tabular}
\caption{Composition of the ice mixture measured with FTIR spectroscopy at 8 K. For each experiment, the first row indicates the abundance 
(by number of molecules) 
in percent relative to water, while the second row shows the absolute abundance (also in percent).}
\begin{footnotesize}
{a Heating rate = 1 K/min\\
b Heating rate = 2 K/min}
\end{footnotesize}
\label{exp}
\end{table}

\section{Experimental results and discussion}

\subsection{Mixture effects on TPD curves of a pre-cometary ice analog}

\begin{figure}
 \includegraphics[width=9.3cm]{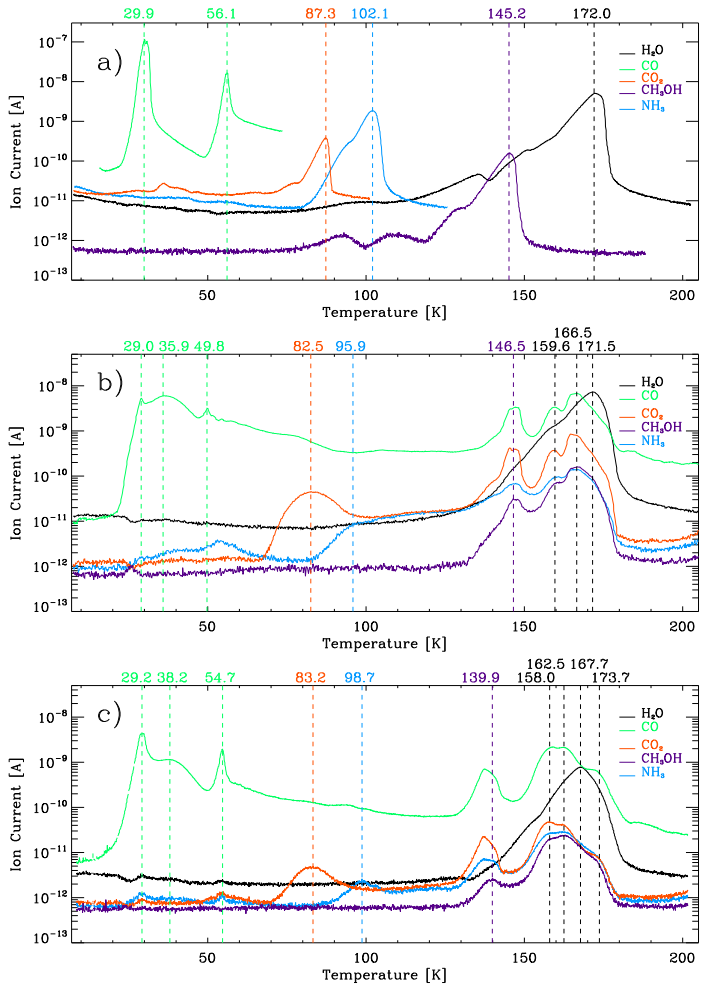}
 \caption{The TPD curves of each component in experiments a) E1-E5 (pure ices), b) E10 (pre-cometary ice mixture with heating rate = 2 K/min).   
 Desorption of a molecule with mass fragment m/z=15 at T $<$ 60 K does not correspond to NH$_{3}$, since that feature is not shared with 
 the m/z=17 fragment (not shown for clarity), and it may be due to a contaminant, c) E11 (pre-cometary ice mixture with heating rate = 1K/min). 
Temperatures of desorption peaks are indicated. 
Ion current in the y-axis corresponds approximately to the partial pressure (mbar) in the main chamber. 
Note that the y-axis is on a logarithmic scale.}
\label{figura1}
\end{figure}

Figure \ref{figura1}a shows the TPD curves of the pure ices (E1-E5), which are measured with the QMS of ISAC. 
The ices were deposited at 8 K and subsequently warmed up with a heating rate of 1 K/min.  
All curves are similar to those reported in previous works (e.g., Collings et al. 2004) with a broad feature that peaks at a different temperature 
according to the sublimation energy or the volatility of the species. 
The bumps that appear in these broad features are due to phase changes occurring in the ice between different amorphous structures, 
or from an amorphous to a crystalline phase.  
In all cases the previous structure has a higher vapor pressure.  
Therefore, the desorption rate first increases with temperature, then decreases as the phase change occurs, and increases again after it is completed. 
These bumps are more easily observed for thick ices (Brown et al. 2006).

Unlike the rest of the studied molecules, CO molecules bind more strongly to the KBr substrate than to themselves. 
Therefore, its curve presents two peaks: the first peak at T = 29.9 K corresponds to the bulk (multilayer) desorption, 
while the second peak at T = 56.1 K corresponds to molecules desorbing directly from the substrate (monolayer desorption).

Figure \ref{figura1}b shows the TPD curve of each ice component in a pre-cometary ice mixture  
co-deposited at 8 K and warmed up with a heating rate of 2 K/min (experiment E10).                                                                    
All species show desorption peaks at temperatures near the ones corresponding to pure ices. 
Desorption of NH$_{3}$ and CH$_{3}$OH at temperatures compatible with 
pure 
desorption was not previously documented for co-deposited mixtures 
(Collings et al. 2004; Brown et al. 2006).   
Desorption of molecules from a pure ice environment 
is considered evidence of segregation occuring in the ice mixture, at least to some extent (Bisschop et al. 
2006). Segregation of species during thermal annealing is studied in more depth in Sect. \ref{ev}.
                       
We detect 
the two desorption peaks of CO between 30 K and 50 K, as reported in Collings et al. (2004) for a co-deposited binary mixture, and 
also 
a third peak at a temperature below 30 K thanks to a better sensitivity of the QMS and a slower heating rate.
The 
peak detected at 29.0 K and the peak at 49.8 K correspond to the multilayer and monolayer desorption of CO ice, respectively, which are also observed in 
the pure ice experiment (E2, Fig. \ref{figura1}a). 
The origin of the peak at 35.9 K, attributed by Collings et al. (2004) to monolayer desorption, is not fully understood. 

In addition, CO molecules in the ice keep 
desorbing continuously at temperatures higher than those corresponding to multilayer and monolayer desorptions, according to the QMS data, see 
Fig. \ref{figura1}a.
This was confirmed by means of IR spectroscopy.
Figure \ref{CO} shows the evolution of the C=O stretching band at 2139 cm$^{-1}$ during warm-up of the pre-cometary ice mixture in experiment E10.  
The absorbance 
decreases between the monolayer desorption at T $\sim$ 50 K and the 
next desorption peak 
at T $\sim$ 146 K.  
Assuming that the band strength does not change very significantly with temperature, 
as it is the case for pure CO ices, 
the column density drops from $\sim$ 81 \% of its initial value at T $\sim$ 63 K to $\sim$ 48 \% at T $\sim$ 133 K.

All the species except H$_{2}$O present desorption peaks 
at temperatures higher than the corresponding temperature for pure ices. 
This indicates that these molecules are 
retained 
in the water ice structure (Bar-Nun et al. 1985; Collings et al. 2004; Fayolle et al. 2011).
A fraction of the trapped molecules is released in the "volcano" desorption (Smith et al. 1997) 
when the change from amorphous to cubic crystalline water ice occurs at T $\sim$ 160 K. 
Volcano desorption of NH$_{3}$ molecules in co-deposited mixtures with water was not reported in Collings et al. (2004)
, while volcano desorption of CH$_{3}$OH had already been detected by Brown et al. (2006). 
Another fraction of the trapped molecules co-desorbs later with water, which peaks at T $\sim$ 171 K.

A second volcano desorption in the 160 - 170 K temperature range, corresponding to the release of molecules during the  
phase change from cubic to hexagonal water ice, is probably responsible for the shift between the co-desorption peaks temperature (T $\sim$ 166 K) and the water desorption peak 
temperature. 
This transformation takes place at 213 K in pure ices (Dowell \& Rinfret, 1960), but the presence of large 
quantities of trapped molecules eases the phase change at a lower temperature range (Notesco \& Bar-Nun, 2000). 
Therefore, we found that
\begin{itemize}
 \item No shift is detected when the concentration of H$_{2}$O is above 85\% (E6, Fig. \ref{figura2}a; E8, Fig. \ref{figura2}d).
\item A shift of $\sim$ 5 K is observed when the concentration is $\sim$ 70\% 
( E7, Fig. \ref{figura2}b; E10, Fig \ref{figura1}b; E11 Fig. \ref{figura1}c).
\item A shift of $\sim$ 10 K is observed when the concentration is below 60\% (E9, Fig. \ref{figura2}e). 
\end{itemize}

The three types of desorption described so far (desorption as for the pure ice, volcano desorption and co-desorption with water) are common to most of 
the binary mixtures with H$_{2}$O. 
In a first approximation, the thermal desorption behavior of interstellar and cometary ices is dominated by interactions with the water ice matrix 
(diffusion and entrapment of molecules within the water ice structure). 
However, 
co-desorption of the three most volatile species (CO, CO$_{2}$, and NH$_{3}$) 
with CH$_{3}$OH at T $\sim$ 146 K
is also observed 
in three of the studied multicomponent mixtures,  
where desorption of methanol at a temperature near the one corresponding to the pure ice is
detected 
(E10, Fig \ref{figura1}b; E11, Fig. \ref{figura1}c; and E7, Fig. \ref{figura2}b).  
These co-desorption peaks cannot be reproduced in binary ice mixtures with water. 

\subsubsection{Effects of the heating rate}
Figure \ref{figura1}c shows the TPD curves of a pre-cometary ice mixture with a heating rate of 1 K/min (E11). 
The curves are rather similar to those observed in experiment E10 (Fig. \ref{figura1}b) with a similar ice composition and thickness and a 
heating rate two times faster (2 K/min). 
All desorption peaks observed in Fig. \ref{figura1}b are detected in Fig. \ref{figura1}c with the addition of a second co-desorption bump at T $\sim$ 174 K. 
Temperatures of most of the desorption peaks are cooler in experiment E11 (see Table \ref{dosrampas}), as it is expected for a slower heating rate. 
It is unclear, though, why monolayer desorption of CO takes place at higher temperatures. \\

\begin{center}
 \begin{table}
  \centering
  \begin{tabular}{|c|c|c|}
  \hline
  Desorption&\multicolumn{2}{c|}{Temperature (K)}\\
  \cline{2-3}
  peak&E11&E10\\
  \hline
\hline
  CO multilayer&29.2&29.0\\
  \hline
  CO&38.2&35.9\\
  \cline{2-3}
  monolayer&54.7&49.8\\
  \hline
  CO$_{2}$&83.2&82.5\\
  \hline
  NH$_{3}$&98.7&$\sim$95.9\\
  \hline
  CH$_{3}$OH&140.0&146.5\\ 
  \hline
  volcano&158.0&159.6\\
  \hline
  co-desorption w/ H$_{2}$O&162.5&166.5\\
  \hline
  H$_{2}$O&167.7&171.5\\
  \hline
  \end{tabular}
  \caption{Temperature of desorption peaks in experiments E11 (Fig. \ref{figura1}c, heating rate = 1 K/min) and E10 (Fig. \ref{figura1}b, heating rate = 2 K/min).
Temperature of the ammonia desorption peak in E10 (Fig. \ref{figura1}b) is not well constrained due to the shape of the curve.
Temperatures of volcano and co-desorption peaks are approximate, since they do not coincide exactly for all the species.}
 \end{table}
\label{dosrampas}
\end{center}

\subsection{Binding energies of pure ices}
The desorption rate of pure ices is given by the Polanyi-Wigner equation, 

\begin{equation}
 \frac{dN_{g}}{dt} = \nu_{i} N_{s}^{i} exp(-E_{des}/T), 
\end{equation}

where $dN_{g}/dt$ is the desorption rate in molecules cm$^{-2}$ s$^{-1}$, 
$\nu_{i}$ a pre-exponential constant for order $i$ in molecules$^{1-i}$ cm$^{-2(1-i)}$ s$^{-1}$, 
$N_{s}$ the column density of the desorbing ice at time $t$ in molecules cm$^{-2}$, 
$E_{des}$ the binding energy in K, 
and $T$ the ice temperature in K at time $t$. 
Both $E_{des}$ and $\nu_{i}$ depend, in principle, on the ice thickness, but this dependence is not thought to be large, even for non-zero desorption orders 
(Bisschop et al. 2006). 
However, Bolina et al. (2005a) reported a difference of $\sim$ 20 \% in the binding energy of CH$_{3}$OH using ices with a range of thicknesses 
spanning one order of magnitude and a non-zero desorption order. 

Desorption peaks of pure ices can thus be fitted using the Polanyi-Wigner expression and the relation, 

\begin{equation}
 \frac{dN_{g}}{dt} = \frac{dN_{g}}{dT} \frac{dT}{dt}, 
\end{equation}

where $dT/dt$ is the heating rate.

Table 
3 
presents an estimate of the pre-exponential constant and the binding energy for the five pure ices, assuming zeroth order kinetics, which is 
a reasonable approximation for multilayer desorption peaks (Noble et al. 2012). 
We used a $\chi^2$ parameter to find the best fit to the experimental desorption peaks. 
Since this parameter is strongly dependent on the range of temperatures for which the fit is done, we have used the binding energy 
estimated by the TPD peak position with equation 4 (Attard \& Barnes 1998) as a reference. 

\begin{equation}
E_{des} = T_{des} \times 30.068.
\end{equation}

Errors indicated in Table 
3 
arise from the range of values leading to a similar $\chi^2$. 
The pre-exponential constant found for the CO desorption peak is one order of magnitude larger than the values previously reported, while the binding energy 
coincides within errors (Mu\~noz Caro et al. 2010; Noble et al. 2012). 
The binding energy obtained for the CO$_{2}$ desorption peak also coincides within errors with the value reported by Sandford \& Allamandola (1990), 
although it is $\sim$ 10 \% larger than that reported by Noble et al. (2012) 
while the pre-exponential constant is a factor $\sim$ 5 larger in our case. 
Bolina et al. (2005b) used a desorption order of 0.25$\pm$0.05 for the NH$_{3}$ desorption peak leading to a binding energy close to the one reported here, but 
a pre-exponential constant two orders of magnitude lower.  
In the case of CH$_{3}$OH, the binding energy in Table 
3 
is within the range of values reported by Bolina et al. (2005a), 
where a desorption order of 0.35$\pm$0.21 is used. 
Pre-exponential constant for H$_{2}$O desorption coincides within errors with that reported in Fraser et al. (2001), although the binding energy we obtained 
is $\sim$ 10 \% lower.

\begin{center}
 \begin{table}
  \centering
  \begin{tabular}{|c|c|c|c|}
  \hline
  &$\nu_{0}$/10$^{27}$&$E_{des}$&Estimated\\
  Species&(mol cm$^{-2}$s$^{-1}$)&(K)&$E_{des}$ (K)\\
  \hline
  CO&6.5$^{+15.3}_{-4.6}$&890$\pm$35&899\\
  \hline
  CO$_{2}$&5.2$^{+4.3}_{-1.6}$&2605$^{+50}_{-30}$&2625\\
  \hline
  NH$_{3}$&2.1$^{+2.7}_{-1.2}$&2965$\pm$80&3070\\
  \hline
  CH$_{3}$OH&5.3$^{+16.0}_{-2.7}$&4355$^{+200}_{-100}$&4366\\
  \hline
  H$_{2}$O&32.1$^{+12.8}_{-8.5}$&5165$\pm$55&5172\\
  \hline
  \end{tabular}
  \caption{Desorption rate parameters for the five pure ices. 
Columns 2 and 3: Pre-exponential constants and binding energies assuming the Polanyi-Wigner desorption rate equation 
and zeroth order kinetics.  Errors arise from the range of values leading to a similar $\chi^2$. Column 4: estimated binding energy from equation 4.}
 \end{table}
\label{pw}
\end{center}

\subsection{Entrapment of molecules in multicomponent ices}
\label{trapp}

Since ice diffusion barriers are proportional to the volatility of the species (\"Oberg et al. 2009), 
volatile molecules like CO or, to a lesser extent, CO$_{2}$, are able to diffuse more easily through the amorphous solid water. 
Therefore, multilayer and/or monolayer desorption peaks are usually detected for these two species. 
In exchange, less volatile components, like NH$_{3}$ and CH$_{3}$OH, do not diffuse so easily, and are retained to a larger extent within the water ice structure.
Desorption of NH$_{3}$ and CH$_{3}$OH prior to the volcano desorption
had not been previously detected in co-deposited mixtures (Collings et al. 2004; Brown et al. 2006). 
This might be due to a lower sensitivity of the mass spectrometers used in previous works and the faster heating rate applied, which hinders the detection 
of these desorptions. 
These peaks have been detected in our pre-cometary ice mixtures (Figures \ref{figura1}b and c) with similar mixture ratios and thicker ices. 

\begin{figure}
\includegraphics[width=9.4cm]{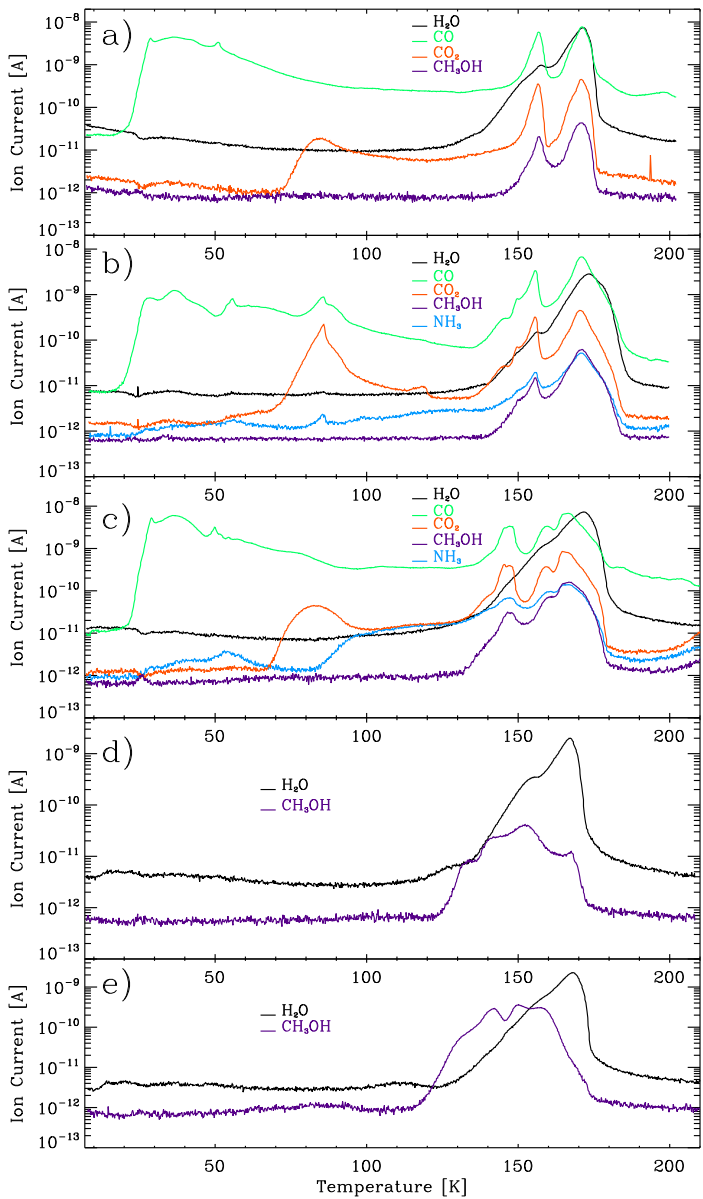}
\caption{The TPD curves of each component in experiments a) E6, b) E7, c) E10, d) E8, and e) E9. 
The concentration of CH$_{3}$OH in the mixture increases from a) to e). 
Ion current in the y-axis corresponds approximately to the partial pressure (mbar) in the main chamber. 
Note that the y-axis is on a logarithmic scale.}
\label{figura2}
\end{figure}

We have studied the effect of the concentration on the entrapment of NH$_{3}$ and CH$_{3}$OH molecules (experiments E6-E10, Fig. \ref{figura2}a-e) 
using ice analogs with a different composition than that of the pre-cometary ice mixtures. 
The fraction of trapped molecules can be inferred from the relative intensities of the peaks corresponding to pure desorption and the volcano and co-desorption 
peaks, which correspond to desorption of trapped molecules. 
As it has been reported for other species, this fraction decreases with an increase of the species concentration relative to water: 

\begin{itemize}
 \item Below a concentration of $\sim$ 2\% relative to water, desorption of CH$_{3}$OH, expected at T $\sim$ 145 K according to Fig. \ref{figura1}a,  
 was not detected (E6, Fig. \ref{figura2}a), and only volcano and co-desorption peaks were observed.   
 Therefore, all CH$_{3}$OH molecules were trapped in the water ice structure.
 \item With a concentration of $\sim$ 3\%, desorption of CH$_{3}$OH at T $\sim$ 149 K, prior to the volcano and co-desorption peaks, 
was clearly detected thanks to the co-desorption of CO and CO$_{2}$ 
(E7, Fig. \ref{figura2}b)). 
The partial pressure of NH$_{3}$ in the chamber also increases slightly in the temperature range at which pure desorption of this species is expected.  
\item Above a concentration of $\sim$ 5\%, desorption peaks equivalent to those of pure NH$_{3}$ and CH$_{3}$OH, respectively, at 95.9 K and 146.5 K 
are clearly identified (E10, Fig. \ref{figura2}c). 
\item The intensity of the desorption peak equivalent to pure CH$_{3}$OH increases with its concentration in the water-dominated ice matrix 
(E8, Fig. \ref{figura2}d; E9, Fig. \ref{figura2}e).
\end{itemize}

Clear trends are observed regardless of the total column density deposited in each 
experiment, which indicates that the total thickness of the ice is not as relevant for ices of a few hundred monolayers as it is for thin ices. 

\subsection{Mixture effects on IR spectra of a pre-cometary ice analog}
\label{IR}

\begin{figure}
\includegraphics[width=9.25cm]{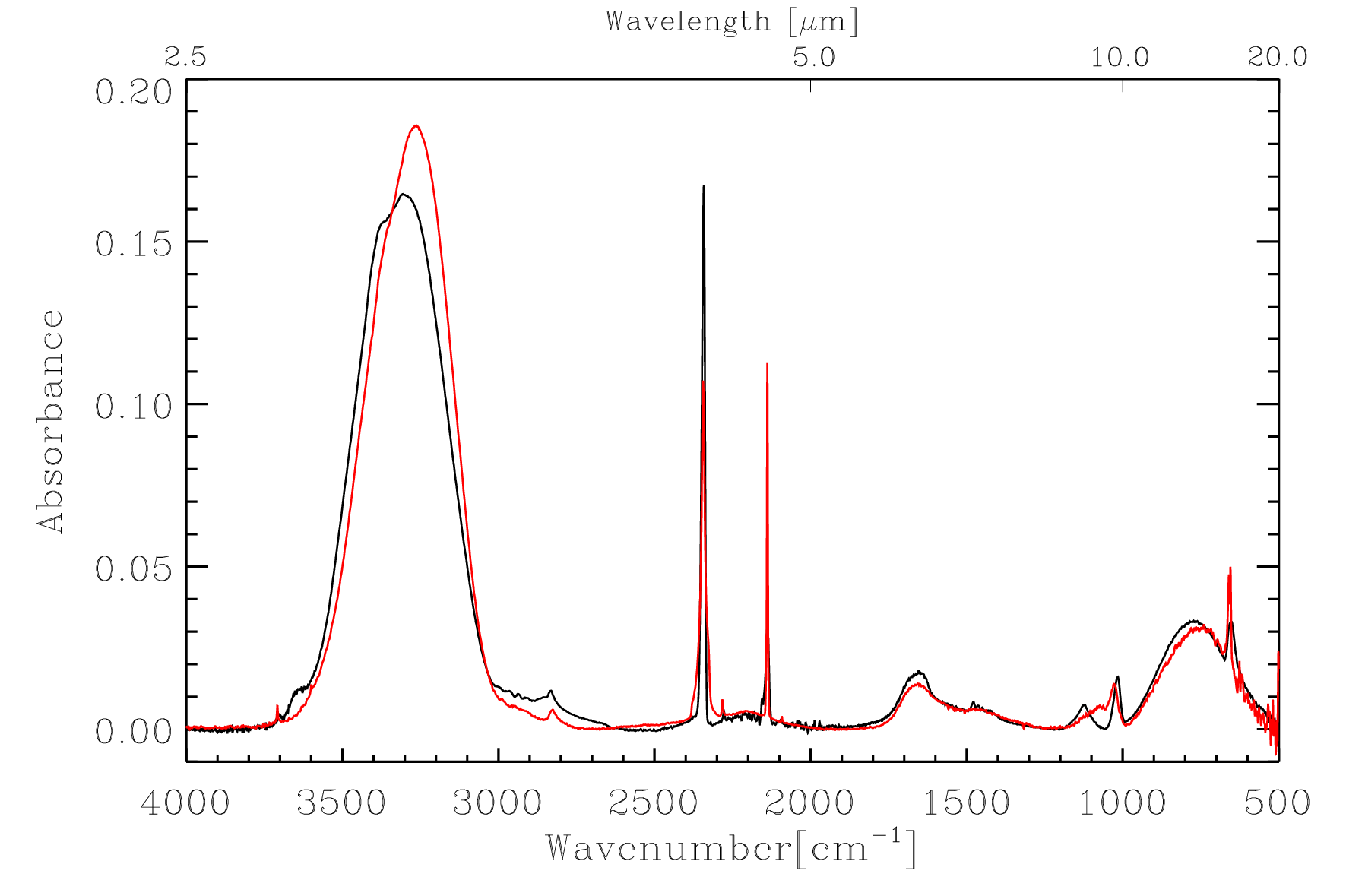}
\caption{Black: IR spectra of a pre-cometary ice mixture (E11) registered at 8 K. Red: Sum of the IR spectra of individual pure ice components (E1-E5) 
registered at 8 K; 
they were multiplied by a factor to reproduce the composition of the pre-cometary ice mixture.}
\label{IRcometaria_1}
\end{figure}

The IR spectra collected during the heating of the ice analogs compliment the analysis of the TPD data. 
Figure \ref{IRcometaria_1} shows the infrared spectrum at 8 K of the pre-cometary ice mixture in experiment E11, along with a spectral fit 
consisting of the sum of the individual pure ice spectra (E1-E5) that are also registered at 8 K, after scaling with a factor 
to reproduce the composition of the ice mixture.

The IR spectrum of the pre-cometary ice mixture appears to be fairly well reproduced by the sum of individual spectra of pure ices,  
although some significant differences are observed, which are discussed below in order of decreasing wavenumber. 
Most of these differences are due to interactions between water molecules and the rest of ice components  
and have been previously reported for binary mixtures. 
This is in line with the similarity between 
TPD curves in pre-cometary ice mixtures and in binary mixtures with the exception for the co-desorption of CO, CO$_{2}$, and NH$_{3}$ with CH$_{3}$OH, as mentioned 
in Sect. 3.1. 

\smallskip

There is a new feature at $\sim$ 3650 cm$^{-1}$ which is not reproduced by the sum of pure ice spectra. 
This feature arises from the interactions of CO and CO$_{2}$ molecules with
 the amorphous solid water (Hagen et al. 1983). 
Hydrogen bonds formed between CO$_{2}$ and H$_{2}$O molecules are weaker than those formed between water molecules. The molecule CO is not even able to form these bonds. 
Therefore, there are some water OH-groups, who are not connected to the hydrogen bonded network  
and whose O-H stretching mode has a higher frequency, leading to the appearance of this feature in the blue side of the O-H stretching band.

Moreover, the extent of intermolecular coupling in the water ice structure becomes smaller and the hydrogen bonds 
that do form become weaker, which explains the blueshift of the O-H stretching band of H$_{2}$O, from 3262 cm$^{-1}$ to 3307 cm$^{-1}$ due to the presence of CO and CO$_{2}$. 
In addition, this band has a lower intensity than expected, since the induced polarization is reduced, 
but has a more intense low-frequency wing due to the presence of CH$_{3}$OH and NH$_{3}$ in the ice mixture (Hagen et al. 1983). 

\smallskip

The band strength of the N-H stretching mode of NH$_{3}$ seems to be also enhanced in the mixture, since the shoulder at $\sim$ 3370 cm$^{-1}$ 
is more intense than expected. 
Due to the overlap with the O-H stretching mode of H$_{2}$O, this mixture effect is very difficult to quantify (d'Hendecourt \& Allamandola 1986).

\smallskip

The C=O stretching band of CO$_{2}$ at 2344 cm$^{-1}$ is slightly redshifted to 2342 cm$^{-1}$ (see left panel of Fig. \ref{CO2com}),  and it is sharper in the mixture. 
The behavior of this band in mixtures with H$_{2}$O depends on the concentration of CO$_{2}$ and also on the presence of other polar molecules, such as 
CH$_{3}$OH 
(Sandford \& Allamandola 1990; Ehrenfreund et al. 1999). 

\smallskip

The C=O stretching band of CO at 2139 cm$^{-1}$ has a shoulder on its blue side, peaking at $\sim$ 2152 cm$^{-1}$ (see Fig. \ref{CO}). 
This shoulder is believed to originate from the interaction of CO molecules with dangling O-H bonds of water molecules 
at a surface or discontinuity 
that cannot complete a tetrahedrally hydrogen bonded network (Collings et al. 2003a). 
Surprisingly, this peak is not observed in astronomical spectra, which can be reproduced with a mixture of solid CO and CH$_{3}$OH without the presence of 
H$_{2}$O (Cuppen et al. 2011).

\smallskip

The band strength of the O-H bending mode at $\sim$ 1650 cm$^{-1}$ is increased in the mixture due to the presence of CO and CO$_{2}$ (Hagen et al. 1983). 

\smallskip

The umbrella mode of NH$_{3}$ at 1067 cm$^{-1}$ is blueshifted to 1124 cm$^{-1}$ in the mixture. 
This is observed in binary mixtures with H$_{2}$O (Hagen et al. 1983; d'Hendecourt \& Allamandola 1986). 
The relatively strong shift of this band manifests the intermolecular bond strength between NH$_{3}$ and H$_{2}$O molecules.

On the other hand, the C-O stretching band of CH$_{3}$OH is redshifted from 1025 cm$^{-1}$ to 1016 cm$^{-1}$. 
This effect has been also observed in binary mixtures (d'Hendecourt \& Allamandola 1986). 
This relatively small shift indicates that the hydrogen bond strength between CH$_{3}$OH and H$_{2}$O molecules in the ice, and the CH$_{3}$OH - CH$_{3}$OH bonds
are only slightly different. This observation is confirmed by the similar binding energies inferred from the desorption temperatures. 
\smallskip

The degenerate bending modes of CO$_{2}$ at $\sim$ 655 cm$^{-1}$ lose the double-peak structure in the mixture, 
as previously reported for binary mixtures with water (Sandford \& Allamandola 1990). 

\subsection{Evolution of IR spectra during thermal annealing}	
\label{ev}

\begin{figure}
\includegraphics[width=9cm]{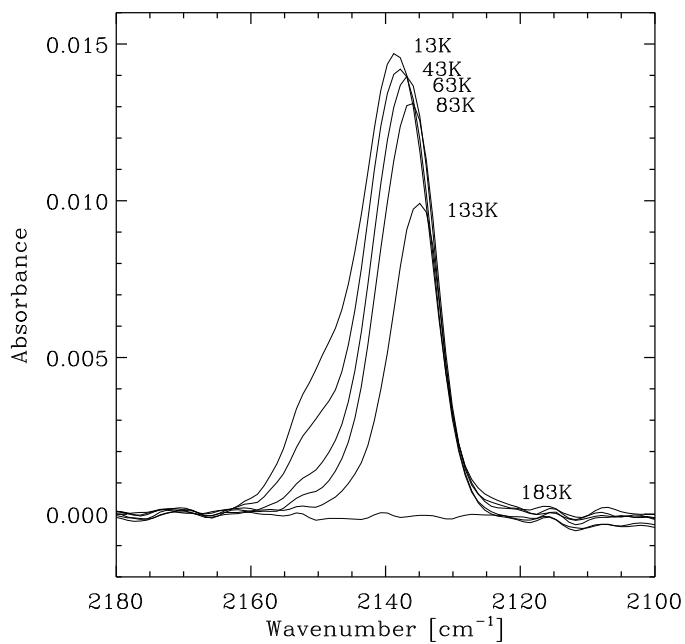}
\caption{Evolution of the C=O stretching band of CO during the warm-up of a pre-cometary ice mixture (experiment E10).} 
\label{CO}
\end{figure}

Figure \ref{CO} shows the gradual disappearance of the shoulder in the blue side of the C=O stretching band of CO during the warm-up. 
It is negligible at $\sim$ 63 K and has completely disappeared at $\sim$ 83 K.
This phenomenon is thought to be an effect of the decrease in the number of dangling O-H bonds in the water ice structure during the transition 
from the highly porous to the less porous amorphous solid water 
(Sandford et al. 1988; Collings et al. 2003b). 
\"Oberg et al. (2009) observed an increase in the intensity of the main peak at 2138 cm$^{-1}$ relative to the shoulder at T $\sim$ 25 K  
(before the onset of the phase transition), suggesting that the change in the IR band is due to segregation of CO in the ice. 
Since the loss of the shoulder in Fig. \ref{CO} takes place along with a redshift of the main peak, which is characteristic of binary mixtures with water (Sandford et al. 
1988) in our experiment these effects are most likely due to structural changes in the ice.
Therefore, no strong evidence of CO segregation is found in the IR spectra, despite the detection of desorption peaks at temperatures 
corresponding to pure CO ice in the TPD curves. 
This probably means that CO segregation takes place only to a small extent (to a small spatial extent within the ice and/or in a short time span before multilayer  
desorption occurs) and its effects cannot be detected with the FTIR spectrometer because it is less sensitive than the QMS. 

\bigskip

\begin{figure}
\includegraphics[width=9.1cm]{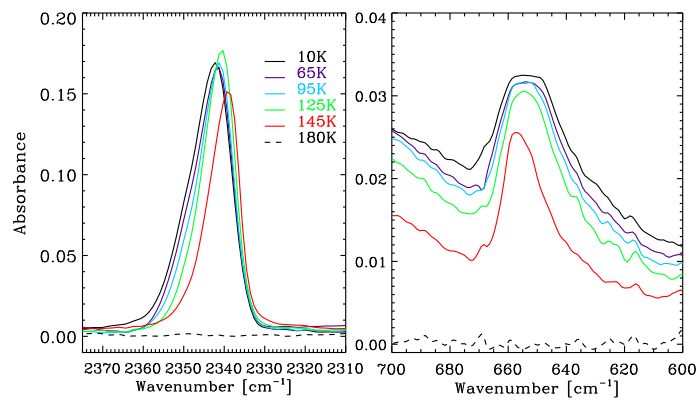}
\caption{Evolution of the IR bands of CO$_{2}$ during the warm-up of a pre-cometary ice mixture (experiment E11). Left: C=O stretching mode at $\sim$ 
2342 cm$^{-1}$. Right: bending mode at $\sim$ 655 cm$^{-1}$.}
\label{CO2com}
\end{figure}

The evolution of the CO$_{2}$ stretching and bending bands is shown in the left and right panel of Fig. \ref{CO2com}, respectively. 
The C=O stretching band at 2342 cm$^{-1}$ slightly increases its intensity between 65 K and 125 K 
at the same temperature range in which pure CO$_{2}$ desorption takes place. 
This behavior is shared with the C=O stretching band of pure CO$_{2}$ ice, but we do not observe other effects of ice segregation, as reported by 
\"Oberg at al. (2009)
for a binary mixture of CO$_{2}$ and H$_{2}$O, 
such as the blueshift or the appearance of a new feature in the blue side of this band (Instead, the redshift increases to 2339 cm$^{-1}$ 
at higher temperatures in our ice mixture.), 
or the double peak structure of the degenerate bending modes.   
Again, there is no strong evidence of CO$_{2}$ segregation in our IR spectra, probably 
because it takes place only to a small extent, as it does in the case of CO molecules. 

\bigskip

\begin{figure}
\includegraphics[width=9.1cm]{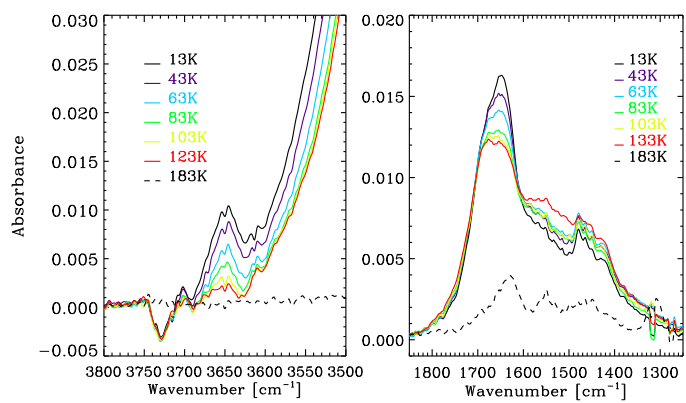}
\caption{Evolution of the IR bands of H$_{2}$O during the warm-up of a pre-cometary ice mixture (experiment E10). Left: O-H stretching mode at $\sim$ 3650 
cm$^{-1}$. 
Negative absorbances are due to the atmospheric compensation applied by the spectrometer software. 
Right: O-H bending mode at $\sim$ 1654 cm$^{-1}$.}
\label{h2ocom}
\end{figure}

The feature at $\sim$3650 cm$^{-1}$, which corresponds to the O-H stretching mode of water molecules that are not completely connected to the hydrogen 
bonded network due to interactions with CO and CO$_{2}$ molecules, 
gradually disappears during the warm-up (left panel of Fig. \ref{h2ocom}). 
This is most likely an effect of transition to a more compact ice structure (Isokoski 2013), 
since there is no reliable evidence of CO and CO$_{2}$ segregation (see above). 
The intensity of the bending mode of H$_{2}$O decreases in the same temperature range (right panel of Fig. \ref{h2ocom}), 
which can also be explained by compaction. 

\bigskip

\begin{figure}
\includegraphics[width=9cm]{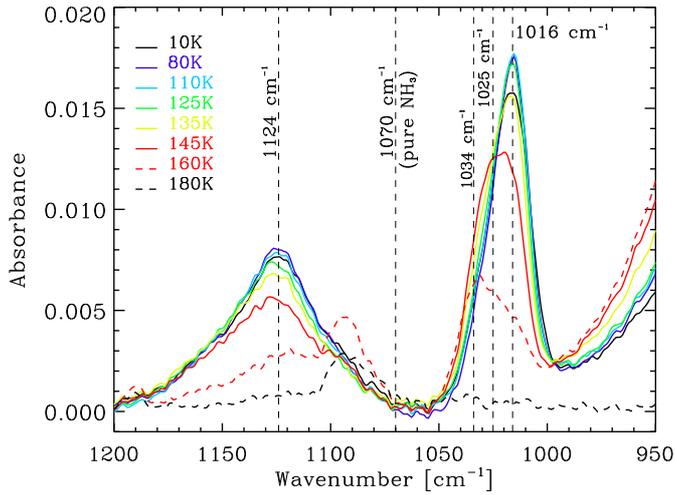}
\caption{Evolution of the umbrella mode of NH$_{3}$ and the C-O stretching mode of CH$_{3}$OH during the warm-up of a pre-cometary ice mixture 
(experiment E11).} 
\label{IR1100}
\end{figure}

Figure \ref{IR1100} shows the evolution of the umbrella mode of NH$_{3}$ at 1124 cm$^{-1}$ and the C-O stretching mode of CH$_{3}$OH at 1016 cm$^{-1}$. 
No substantial changes are observed in the former, but the latter changes its band profile and peak frequency during the warm-up. 
Above a temperature of T $\sim$ 110 K before the onset of the desorption of CH$_{3}$OH, the peak shifts toward shorter wavelengths. 
At  T $\sim$ 145 K, the band is broader and peaks at a frequency close to that of pure CH$_{3}$OH ice. At higher temperatures, the red component of the band 
gradually disappears and the peak is shifted to $\sim$ 1034 cm$^{-1}$.  
This indicates that a fraction of the CH$_{3}$OH molecules segregates  
in the ice at temperatures above T $\sim$ 110 K, while other CH$_{3}$OH molecules form a type II clathrate hydrate (Blake et al. 1991). 
Segregated molecules are able to desorb more easily than molecules in the clathrate hydrates, which mainly desorb at higher temperatures.

\subsubsection{CH$_{3}$OH segregation}

The CH$_{3}$OH molecules segregate enough to an extent to be detected by the FTIR spectrometer. 
We have fitted the C-O stretching band of CH$_{3}$OH in the IR spectra collected above T $\sim$ 110 K 
by using three Gaussians to estimate the contribution of the three types of methanol that can be found in the ice sample, 
namely: mixed methanol (centered at 1016 cm$^{-1}$), segregated methanol (centered at 1025 cm$^{-1}$), and methanol forming a type II clathrate hydrate 
with water molecules (centered at 1034 cm$^{-1}$). 
A fourth Gaussian was used to take into account the contribution of the blue wing of the H$_{2}$O librational band. 

\begin{center}
\begin{figure}
\includegraphics[width=9.25cm]{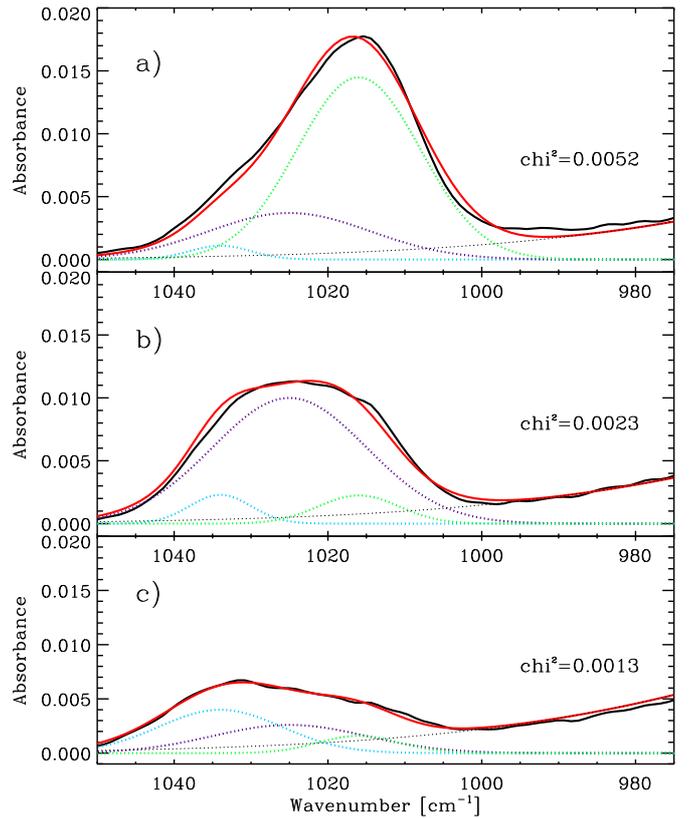}
\caption{The C-O stretching band of CH$_{3}$OH (black) fitted with the sum of three Gaussians (red), which account for: 
mixed methanol (green dotted, centered at 1016 cm$^{-1}$), 
segregated methanol (purple dotted, centered at 1025 cm$^{-1}$), and methanol forming a type II clathrate hydrate with water molecules (blue dotted, 
centered at 1034 cm$^{-1}$). A fourth Gaussian has been used to take into account the contribution of the blue wing of the H$_{2}$O librational band 
(black dotted). The IR spectra correspond to a pre-cometary ice analog (E11) at a) T = 110 K, b) T = 150 K, and c) T = 160 K. 
The $\chi^{2}$ parameter is indicated for each fit.}  
\label{gauss}
\end{figure}
\end{center}

\begin{figure}
\includegraphics[width=9.25cm]{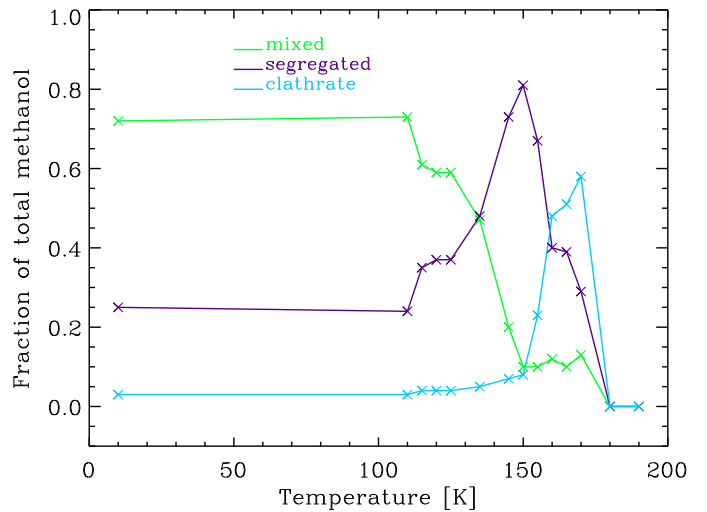}
\caption{Evolution of the contribution of mixed methanol (green), segregated methanol (purple), and methanol forming a type II clathrate hydrate (blue) to 
the total area of the C-O stretching band of CH$_{3}$OH during warm-up of a pre-cometary ice mixture (E11), according to fit with three Gaussians of the IR band. 
Temperatures for which the fit was done are indicated. 
Figure \ref{gauss} shows three selected fits as an example.}
\label{areas}
\end{figure}

Figure \ref{gauss} shows the fit for three spectra at T $\sim$ 110 K, 150 K, and 160 K, respectively. 
We used a $\chi^{2}$ parameter to find the best fit in each case. 
The evolution of the contribution of each Gaussian to the total area of the C-O stretching band is shown in Fig. \ref{areas}. 
Above T $\sim$ 110 K, the contribution of segregated methanol and methanol in a type II clathrate hydrate to a lesser extent increases, 
while that of the mixed methanol decreases. 
At T $\sim$ 150 K, before the volcano desorption occurs, segregated methanol dominates in the ice mixture 
and above that temperature methanol in a type II clathrate hydrate becomes the major contributor to the IR band until it is completely desorbed.

\bigskip

\begin{figure}
\includegraphics[width=9.25cm]{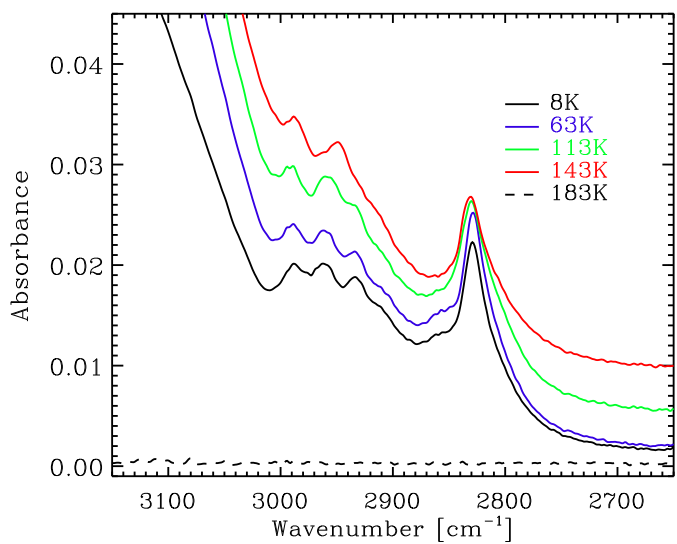}
\caption{Evolution of the C-H stretching modes of CH$_{3}$OH during the warm-up of a binary mixture (experiment E9). 
Spectra are offset for clarity.}
\label{IR3000}
\end{figure}

A final confirmation of the methanol segregation can be found in the C-H stretching region of the spectrum.
At these high temperatures, pure (segregated) CH$_{3}$OH should be in its crystalline form. 
When pure CH$_{3}$OH ice changes from the amorphous to the crystalline $\alpha$ structure, the asymmetric C-H stretching mode 
(which at 8 K is a wide band peaking at 2930 cm$^{-1}$ with a shoulder at $\sim$ 2984 cm$^{-1}$)   
is perfectly resolved into the a' ($\nu_{2}$) and a'' ($\nu_{9}$) modes at 2984 cm$^{-1}$ and 2856 cm$^{-1}$, respectively, 
while the symmetric mode ($\nu_{3}$) at 2827 cm$^{-1}$ 
remains basically unchanged (Falk \& Whalley 1961). 
Unfortunately, the $\nu_{9}$ mode is barely detected in the pre-cometary ice mixture (Fig. \ref{IRcometaria_1}), 
and it is not possible to confirm this process, 
but it is observed in a binary mixture with 
a higher concentration of methanol (experiment E9, Fig. \ref{IR3000}). 
In this mixture, we observe three peaks at 2989 cm$^{-1}$, 2960 cm$^{-1}$, and 2933 cm$^{-1}$ at 8 K, instead of the wide band peaking at 2930 cm$^{-1}$. 
These peaks may arise from the interaction between CH$_{3}$OH and H$_{2}$O. 
When the temperature raises to T $\sim$ 113 K, these three peaks begin to change, and at T $\sim$ 143 K, two peaks  
are clearly identified at frequencies near those of the a' and a'' asymmetric stretching modes (2988 cm$^{-1}$ and 2948 cm$^{-1}$).
This proves that a fraction of CH$_{3}$OH molecules is actually segregating into its crystalline form.

It has been pointed out previously that segregation takes place if diffusion of molecules in the ice is possible and it is energetically favorable 
for molecules of the same species to group together, being ice diffusion barriers proportional to their volatility (\"Oberg et al. 2009). 
Therefore, the most volatile ice species, like CO, can diffuse easily within the water ice structure and, eventually, meet other CO molecules. 
However, since the CO - CO interaction is the weakest among the species studied here, 
only small groups of segregated CO molecules can be formed before they reach the ice surface and 
desorb or, alternatively, the segregated-CO seed is torn apart by stronger interactions with other species (mainly H$_{2}$O, since it is the major constituent 
in our pre-cometary ice analogs). 
On the other hand, less volatile species like CH$_{3}$OH can form larger groups of segregated molecules when the ice temperature is sufficiently high, 
despite the fact that they do not diffuse so easily through the ice structure.
Once methanol molecules meet, the CH$_{3}$OH - CH$_{3}$OH interactions are stronger, since these molecules form hydrogen bonds 
with a similar strength to those formed by H$_{2}$O molecules (Bakkas et al. 1993; note that, unlike CO, CO$_{2}$, and NH$_{3}$, water and methanol are 
both liquids at standard conditions) 
In addition, methanol desorbs at higher temperatures compared to CO, CO$_{2}$, and NH$_{3}$. 
Therefore, segregated-CH$_{3}$OH seeds can grow without desorbing or being disrupted, 
forming larger regions of segregated CH$_{3}$OH which can be detected by the FTIR spectrometer.

\section{Astrophysical implications}
\label{imp}
The heating rate used in our experiments is fast compared to most astrophysical scenarios, as it is generally the case in experimental 
simulations of astrophysical processes. 
The desorption behavior of an ice mixture does not strongly depend on the heating rate (Section 3.1.1), 
but 
differences in the desorption temperatures are 
usually found (e.g., Collings et al. 2004).

Using the kinetic parameters for pure H$_{2}$O ice desorption, Collings et al. (2004) performed a series of theoretical simulations of water ice desorption with 
different heating rates. 
The desorption peak reaches its maximum at T $\sim$ 160 K for a heating rate of 4.8 K/min, similar to the value used in our experiments and at T $\sim$ 105 K 
for a heating rate of 1 K/century, close to the heating rate in hot cores. 

Thermal annealing of interstellar, circumstellar, and cometary ices takes place in three different stages of stellar evolution (see Section 1): 

\begin{itemize}
\item Interstellar ice mantles covering dust grains are formed in dense molecular clouds (T $\sim$ 10 K). These dust grains cycle between the dense 
and the diffuse interstellar medium. Ice mantles do not survive the harsh radiation conditions of the diffuse medium. 
These cycles proceed until dust grains are incorporated in star-forming regions. 
\item Circumstellar ices are heated during the star formation, as the temperature of the central object increases. 
Some grains undergo transient episodes of "flash-heating" lasting up to hundreds of hours according to the fluctuating X-wind model for CAIs and chondrules 
formation 
(Shu et al. 1996, 1997, 2001). 
\item Cometary ices are heated during late thermal evolution of comets around already formed stars, as the comet comes closer to the star during its orbit.
\end{itemize}

As it was mentioned in Section 2, thermal annealing of thick ice mantles, which are formed in outer regions of disks around high and low-mass protostars 
by grain agglomeration,
is mimicked more faithfully in the TPD experiments 
presented in this work 
, due to the thickness of our ice analogs. 
We explain in Section \ref{cometa} that our results can be also applied with some limitations to cometary ices.

\subsection{Circumstellar ices} 

Chemical models of hot cores use the experimental data on TPD of ice analogs 
in the study of their astrochemical network 
(e.g., Viti et al. 2004; Wakelam et al. 2004). 
It was suggested that binary mixtures can be used as templates for more complex ice mixtures to model the desorption of interstellar and circumstellar ices 
(Fayolle et al. 2011). 
In our experiments, we have observed co-desorption of CO, CO$_{2}$, and NH$_{3}$ with CH$_{3}$OH in all the studied mixtures with an abundance of these species 
relative to water above $\sim$ 3\% (which falls in the range of typical abundances found in most astrophysical environments; see Mumma \& Charnley 2011 and 
ref. therein). 
Appropiate TPD experiments with complex (at least tertiary) mixtures are thus needed to incorporate this co-desorption into the models. 

\bigskip

\begin{figure}
\includegraphics[width=9.0cm]{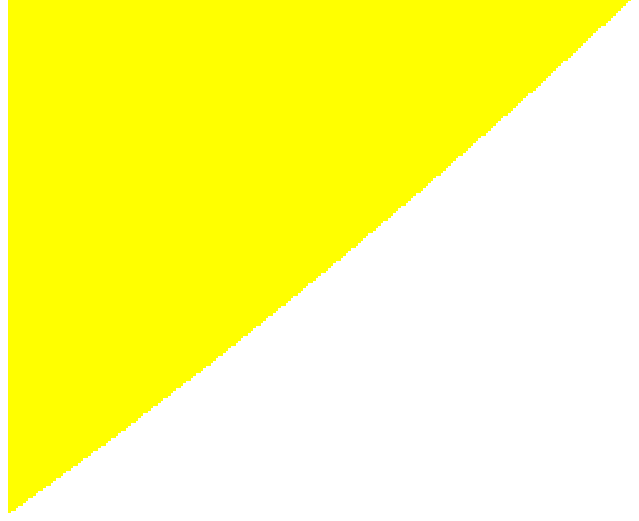}
\caption{
Schematic representation of thermal annealing process of ice mantles in circumstellar regions 
based on our experimental results for the warm-up of a pre-cometary ice mixture (H$_{2}$O:CO:CO$_{2}$:CH$_{3}$OH:NH$_{3}$ = 100:10-20:10-20:5:5). 
Red lines represent the temperature ranges at which the species used in our pre-cometary ice analogs desorb. Approximate positions of desorption peaks are 
indicated with vertical red lines. 
Desorption of NH$_{3}$ and CH$_{3}$OH prior to volcano desorption takes place when their abundance relative to water is above $\sim$ 3 \% in the ice. 
Blue lines represent the temperature ranges at which structural changes in the ice occur. 
The top temperature scale corresponds to a heating rate of 1-2 K/min 
(used in experiments E10 and E11 with our pre-cometary ice mixtures and also suited to the "flash-heating" 
of dust grains during the periodic radial excursions considered in the fluctuating X-wind model for CAIs and chondrules formation). 
The bottom temperature scale corresponds to a first approximation to grains at a certain distance heated with a heating rate of 1 K/century, see text for 
details.}
\label{dibujo2}
\end{figure}

The TPD curves and IR spectral evolution of our pre-cometary ice analogs allowed us to build a schematic representation of the thermal 
annealing process of ice mantles in circumstellar regions (Fig. \ref{dibujo2}). 
The heating rate applied to the pre-cometary ice mixtures in experiments E10 and E11 is of the same order of 
magnitude than that applied to ice mantles during the "flash-heating", as explained by the fluctuating X-wind model for CAIs and chondrules formation. 
Therefore, similar desorption temperatures are expected in this case. 
We have scaled laboratory temperatures to the scenario in which 
grains at a certain distance are heated as the temperature of the central object increases by multiplying them by a factor of 0.62. 
This factor corresponds to the relation between the desorption peak temperature of water ice in our experiments ($\sim$ 170 K) and the one expected  
for a slower heating rate (1 K/century; Collings et al. 2004).    

The temperature gradient generated in the circumstellar envelope by the central protostar leads to chemical segregation in the gas phase. 
High-angular resolution observations of protostellar envelopes represent the best chance to observe this phenomenon in space.

\subsection{Cometary ices}
\label{cometa}

The large size of a comet nucleus (up to 40 km in diameter) establishes several important differences between the simulations 
we have performed and the real processes that take place during the thermal annealing of a comet:

\begin{itemize}
\item Comet nuclei have a temperature gradient in their interiors and the temperature profile is not symmetric (Prialnik et al. 2008). 
The whole ice sample in TPD experiments is kept at the same temperature. 
\item The Deep Impact mission revealed that comet 9P/Temple 1 has an inhomogeneous composition at least at the surface layers (Thomas et al. 2007), while 
our ice analogs are expected to be homogeneously mixed. 
\item Molecules released from the ice can flow both outward (toward the cometary surface
and the coma), 
or inward, 
due to the dynamic percolation through the comet nucleus 
(Notesco \& Bar-Nun 2000). 
\item The energy balance in a comet, dominated by the solar energy during the late thermal evolution of these objects, 
could be more complex than that of our experimental set-up with a small substrate and a controlled warm-up process. 
\end{itemize}

In any case, this work 
should contribute to 
a better understanding of the late thermal cometary evolution, particularly, of 
the formation of the coma,  
as volatiles begin to sublime when the comet approaches the Sun during its orbit.
Assuming a symmetrical temperature profile and a homogeneous ice composition in the nucleus, 
our simulations would correspond to the warm-up process suffered by a finite 
cometary slice, which is not at the same temperature than the rest of the comet nucleus during the heating process. 
The thermal annealing of a comet could thus be seen in a first approximation, as 
the sum of the warm-up of every cometary slice with a missmatch between the temperature of each slice in a certain moment. 
Molecules flowing inward instead of outward could alter the composition of the coma compared to that of the nucleus.

From our results, we can infer that 
when a comet is at a distance to the Sun close enough to heat the cometary surface to a temperature of around 20 K, CO ice starts to desorb  
along with other volatile species that possess a similar desorption temperature, such as N$_{2}$ and O$_{2}$ (Collings et al. 2004), 
 and the coma begins to form. As we have observed in our experiments, once the desorption of CO has begun, it desorbs continuously during the warm-up.

As the comet gets closer to the Sun, less volatile species like CO$_{2}$ desorb successively, although in small amounts. 
Desorption of NH$_{3}$ and CH$_{3}$OH will take place if their abundance relative to water is above $\sim$ 3 \% in the cometary ice.  
When the temperature in the most external layer is high enough to allow desorption of, for instance, CO$_{2}$, desorption of more volatile species 
like CO should be taking place in inner layers due to the temperature gradient of the comet nucleus. 
Similarly, if desorption of NH$_{3}$ begins in the most external ice layer, 
CO$_{2}$ and CO should be desorbing from inner layers.
Therefore, the possible co-desorption of several species with CH$_{3}$OH should have no strong consequences on the composition of the coma  
because the same species that would co-desorb with CH$_{3}$OH should be also desorbing from inner layers. 

The initially (small) CO-rich coma thus becomes more chemically complex as it is enriched with other less volatile molecules during the orbit of the comet. 
The contribution of the inner ice layers makes it difficult to predict the proportion in which volatiles are present in the coma at any time  
and also how this proportion varies as volatiles desorb from deeper inner layers simultaneously to the desorption of less volatile species from 
the outer layers. 
In any case, CO is expected to be the main component in the early stages of activity.

Once the temperature has reached the value at which phase transition between amorphous and crystalline water ice takes place, 
volcano desorption of all volatiles would increase the activity of the comet. 
After this event, desorption of water begins to be quite important, becoming the main component in the coma, 
as it is observed in infrared and radio spectroscopic surveys (Mumma \& Charnley 2011 and ref. therein). 
We note that the detection of species in the coma also includes the "daughters" of parent ice molecules in the nucleus, making the interpretation of cometary 
data even more complex than outlined above. 

\subsection{The ESA-Rosetta cometary mission}

Rosetta is scheduled to arrive at comet 67P/Churyumov-Gerasimenko in July 2014. 
After that, the Rosetta orbiter will remain in close proximity to the comet nucleus and a small lander will be released onto its surface. 
The Rosetta orbiter includes eleven scientific instruments. 
The instrument ROSINA (Rosetta Orbiter Spectrometer for Ion and Neutral Analysis) will determine the composition of the comet atmosphere, playing a similar role 
to the QMS in our experiments. 
The Rosetta lander is known as Philae. There are two evolved gas analyzers (EGAs) among the ten instruments on board Philae: 
Ptolomey and COSAC (COmetary SAmpling and Composition experiment). 
The instrument COSAC is a combined gas chromatograph - mass spectrometer that will analyze the volatile fraction of surface and subsurface samples with special 
attention to the detection and identification of complex organic molecules. Alternatively, it can be operated in sniffing mode where only the mass spectrometer 
is used, analyzing the gas present in the ionization chamber  
(This gas is mainly composed of external gas in the case of high pressures; otherwise internal contamination dominates. Goesmann et al. 2012). 
While the Rosetta orbiter will probably survive perihelion passage, Philae may suffer overheating and cease functioning once the comet approaches its closest 
distance to the Sun.

\bigskip

Using a simple planetary energy balance equation, 
it is possible to estimate a mean temperature of the comet on its surface by knowing its heliocentric distance by the time Rosetta starts the measurements 
and to make bold predictions of what the mass spectrometers will detect based on our laboratory results. 
Assuming that the comet is in radiative equilibrium, meaning that the incoming radiative energy emitted by the Sun and absorbed by the comet is equal to the radiation 
emitted by the comet, the temperature of its surface can be calculated as

\begin{equation}
T=(\frac{L_{sun}(1-a)}{16\pi \sigma r^{2}})^{1/4}, 
\end{equation}

where $L_{sun}$ is the Sun luminosity, $a$ the albedo of the comet, $\sigma$ the Boltzmann constant, and $r$ the heliocentric distance of the comet. 

\begin{center}
\begin{table}
\begin{tabular}{|c|c|c|c|}
\hline
Date&r$_{h}$ (UA)&T (K)&Heating rate (K/min)\\
\hline
01-07-2014&3.80&141.0&6.7$\times 10^{-5}$\\
01-08-2014&3.63&144.0&($\sim$ 3.0 K/month)\\
\hline
01-07-2015&1.35&236.7&2.0$\times 10^{-4}$\\
01-08-2015&1.25&245.7&($\sim$ 0.3 K/day)\\ 
\hline
\end{tabular}
\caption{Estimation of heliocentric distance, temperature, and heating rate of comet 67P/Churyumov-Gerasimenko during its orbit, as calculated using orbital 
parameters provided online by the Jet Propulsion Laboratory (NASA) and equation 5.}
\label{t}
\end{table}
\end{center}

On July 1, 2014, the distance from the comet to the Sun will be at about 3.80 AU. Assuming a cometary albedo of $a = 0.054$ (Kelley et al. 2009), 
the mean temperature at the surface of the comet will be T $\sim$ 141.0 K. 
In our experiments with a heating rate of 1 K/min, water does not desorb substantially at this temperature. 
The coma would thus be composed of small amounts of more volatile species (mostly CO, with the presence of less volatile species like CO$_{2}$ and, maybe, 
NH$_{3}$ and CH$_{3}$OH). 
However, the heating rate applied to the comet during its orbit around the Sun is much slower than the one applied to the ice analogs in our simulations. 
Data from Table \ref{t} allows us to make a rough estimate of the heating rate during the Rosetta approach and near the perihelion after a year. 
As the comet gets closer to the perihelion of its orbit, its velocity, and therefore, also the heating rate, increase. 
Combining the theoretical simulations performed by Collings et al. (2004) of water ice desorption with different heating rates 
and the cometary heating rate estimated in Table \ref{t} by the time of arrival scheduled for Rosetta, we find that
the desorption peak temperature for water ice in comet 67P/Churyumov-Gerasimenko should fall between 115 K and 130 K, approximately, 
This temperatures correspond to a heliocentric distance between 5.75 AU and 4.50 AU, according to equation (5). 
Therefore, superficial water ice should be detected already during the first measurements with an increasing activity in the next months. 
It is difficult, though, to predict how the composition of the coma will be affected by desorption from different layers as the comet is heated by the Sun. 

Observations carried out with the Spitzer Space Telescope during the last passage of comet 67P/Churyumov-Gerasimenko did not detect signs of activity before 
r$_{h}$ = 4.30 AU (Kelley et al. 2009). 
The authors thus expected to find an inactive or weakly active nucleus at the arrival of Rosetta. 
However, ground-based observations carried out at the VLT have recently revealed that the comet presented detectable activity at a heliocentric 
distance of r$_{h}$ = 4.30 AU, which is based on the excess flux in the R-band assigned to dust brightness.  
Therefore, the comet is expected to be active again in March 2014 before the arrival of Rosetta (Snodgrass et al. 2013),   
which is in line with our predictions presented above.
Analysis of previous observations also indicates that the peak of activity takes place about two months after perihelion (de Almeida et al. 2009). 

In addition, 
we expect to use
the experimental data reported here as a benchmark 
for the analysis of  
the data collected by the mass spectrometers on board Rosetta 
as comet 67P/Churyumov-Gerasimenko is heated during its approach to the Sun.   
This will shed light on the relation among the nucleus and the coma compositions, and the ice conditions.  
Despite the differences between these simulations and the real scenario (see Sect. \ref{cometa}), our pre-cometary ice analogs with up to five components 
and astrophysically relevant compositions 
represent the most realistic approach to the thermal annealing of a cometary ice reported to date.

\section{Conclusions}

We have studied the temperature programmed desorption (TPD) of pre-cometary ice analogs with up to five components 
(H$_{2}$O, CO, CO$_{2}$, CH$_{3}$OH, and NH$_{3}$) 
using mass spectrometry to detect molecules desorbing to the gas phase and complementary infrared spectroscopy 
to study the composition, structure, and evolution of the ice during warm-up. 
Comparison with pure ice experiments, which are made of a single molecular component, 
has revealed several 
effects on both the TPD curves and the IR spectra
arising mainly from the interaction between H$_{2}$O, the major constituent of our pre-cometary ice analogs, and the rest of the species, 
and also between CH$_{3}$OH and the most volatile components.

\bigskip

All species present desorption peaks at temperatures near the ones corresponding to pure ices.
For NH$_{3}$ and CH$_{3}$OH,  
these desorption peaks are observed for the first time in water-rich ice analogs at  T $\sim$ 97 K and 
T between 140 -147 K, respectively. 
Both species were previously classified as "water-like", since only peaks at temperatures close to water desorption were reported for co-deposited binary 
mixtures. 
Thermal desorption of ice mixture analogs with different abundances of NH$_{3}$ and CH$_{3}$OH have revealed that  
desorption in our experiments as in the pure ices is detected for ices with an abundance above $\sim$ 3 \% relative to water. 

Desorption of molecules from a pure ice environment suggest segregation of the species in the ice mixture to some extent.  
Only segregation of CH$_{3}$OH molecules was confirmed by means of IR spectroscopy. 
The C-O stretching band of CH$_{3}$OH at $\sim$ 1025 cm$^{-1}$ is redshifted in the pre-cometary ice mixture at 8K, 
but a blueshift is observed at temperatures above $\sim$ 110 K, indicating segregation of this species and formation of a type II clathrate hydrate. 
Segregation of CO, CO$_{2}$, and NH$_{3}$ molecules may occur only to a small extent due to their weaker interactions.

Desorption peaks at temperatures higher than the corresponding temperature for pure ices are also observed for all the species, except for H$_{2}$O, 
which indicates entrapment of molecules within the water ice structure.  
The desorption peaks at T $\sim$ 159 K correspond to volcano desorption after crystallization of the water ice structure, 
while the peaks at T $\sim$ 165 K result from co-desorption with water. 
The shift between the water desorption peak at T $\sim$ 170 K  and the co-desorption peaks temperature 
can be explained by the contribution of a second volcano desorption peak in the 160 - 170 K temperature range 
due to the phase change from cubic to hexagonal water ice. This phase change is eased by the presence of large quantities of trapped molecules 
(Notesco \& Bar-Nun, 2000). 

Prior to the volcano desorption, co-desorption of the more volatile species (CO, CO$_{2}$, and NH$_{3}$) with CH$_{3}$OH has been detected at 
T between 140 -147 K. 

During thermal annealing of the ice analogs, 
CO, CO$_{2}$, and NH$_{3}$ molecules can find themselves in ice regions 
where segregated methanol is the dominant component 
due to the further segregation of CH$_{3}$OH. 
Methanol behaves similarly to water, forming hydrogen bonds with a similar strength. 
Therefore, the three most volatile species co-desorb with methanol and later with water. 
Peaks corresponding to co-desorption of CO, CO$_{2}$, and NH$_{3}$ with methanol cannot be reproduced in experiments with binary mixtures, 
which are extensively used in the study of thermal desorption of interstellar ices. 
In addition, IR spectra confirmed that CO is continuously desorbing between the desorption peaks corresponding to monolayer desorption and co-desorption with 
methanol.

\bigskip

The experiments reported here help to understand the thermal annealing process of circumstellar ices inside hot cores. %Sect. 4.1.
The heating rate applied to our ice analogs is on the same order of magnitude than the one dust grains experience during the periodic radial excursions 
considered in the fluctuating X-wind model for CAIs and chondrules formation (Shu et al. 1996, 1997, 2001). 
A similar behavior is expected for ice mantles at a given distance from the central object, as they are heated with an approximate heating rate of 1 K/century. 
However, differences in desorption peak temperatures are expected. 
Observations with sufficient spatial resolution will serve to test the model sketched in Fig. 4.2. 

With some limitations, these results can be also applied to cometary ices. 
A possible scenario of ice evaporation during comet approach to the Sun is provided in Section 4.2. 
The TPD curves of a pre-cometary ice analog indicate that water could be already detected in the coma of comet 67P/Churyumov-Gerasimenko  
during the first approach of the ESA-Rosetta cometary mission. %Sect. 4.3.

\begin{acknowledgements}
We are grateful to Gustavo A. Cruz D\'iaz, Barbara Michela Giuliano, and Antonio Jim\'enez Escobar for their support on the experiments.
This research was financed by the Spanish MINECO under projects AYA2011-29375 and CONSOLIDER grant CSD2009-00038. 
R. M. D. benefited from a FPI grant from Spanish MINECO.
\end{acknowledgements}

\section*{Bibliography}
Attard, G. \& Barnes, C. 1998, Surfaces (Oxford Science Publications), 72\\
Bar-Nun, A., Herman, G., \& Laufer, D. 1985, Icarus, 63, 317\\
Bakkas, N., Bouteiller, Y., Louteiller, A., Perchard, J.P., \& Racine, S. 1993, J. Chem. Phys. 99, 3335\\
Bisschop, S.E., Fraser, H.J., \"Oberg, K.I., van Dishoeck, E.F., \& Schlemmer, S. 2006, A\&A, 449, 1297\\
Blake, D., Allamandola, L., Sandford, S., Hudgins, D., \& Freund, F. 1991, Science, 254, 548\\
Bolina, A.S., Wolf, A.J., \& Brown, W.A. 2005a, J. Chem. Phys., 122, 4713\\ 
Bolina, A.S., Wolf, A.J., \& Brown, W.A. 2005b,, Sur. Sci. 598, 45
Brown, W.A., Viti., S., Wolff, A.J., \& Bolina, A.S. 2006, Faraday Discussions, 133, 113\\
Collings, M.P., Dever, J.W., Fraser, H.J., \& McCoustra, M.R. 2003a, ApJ, 583, 1058\\
Collings, M.P., Dever, J.W., Fraser, H.J., \& McCoustra, M.R. 2003b, Ap\&SS, 385, 633\\
Collings, M.P., Anderson, M.A., Chen, R., et al. 2004, MNRAS, 354, 1133\\
Cuppen, H.M., Penteado, E.M., Isokoski, K., van der Marel, N., \& Linnartz, H. 2011, MNRAS, 417, 2809\\
de Almeida, A.A., Sanzovo, D.T., Sanzovo, G.C., Boczko, R., \& Torres, R.M. 2009, Adv. Space Res. 43, 1993\\
d'Hendecourt, L.B. \& Allamandola, L.J. 1986, A\&ASS, 64, 453\\
Dowell, L.G. \& Rinfert, A.P. 1960, Nature, 188, 1144\\
Eberhardt, P. 1999, Space Sci. Rev., 90, 45\\
Ehrenfreund, P., Kerkhof, O., Schutte, W.A., et al. 1999, A\&A, 350, 240\\
Falk, M. \& Whalley, E. 1961, J. Chem. Phys. 34, 1554\\
Fayolle, E.C., \"Oberg, K.I., Cuppen, H.M., Visser, R., \& Linnartz, H. 2011, ApJ, 529, A74\\
Fraser, H.J., Collings, M.P., McCoustra, M.R.S., \& Williams, D.A. 2001, MNRAS, 327, 1165\\
Goesmann, F., McKenna-Lawlor, S., Roll, R., et al. 2012, Planet. Space Sci., 66, 187\\
Hagen, W. 1981, Chem. Phys., 56, 367\\
Hagen, W., Tielens, A.G.G.M., \& Greenberg, J.M. 1983, A\&ASS, 51, 389\\
Herbst, E. \& van Dischoeck, E.F. 2009, Annu. Rev. Astron. Astrophys., 47, 427\\
Isokoski, K. 2013, PhD Thesis\\
Jiang, G.J., Person, W.B., \& Brown, K.G. 1975, J. Chem. Phys., 62, 1201\\
Kelley, M.S., Wooden, D.H., Tubiana, C., et al. 2009, AJ, 137, 4633\\
Mumma, M.J. \& Charnley, S.B. 2011, Annu. Rev. Astron. Astrophys., 49, 471\\
Mu\~noz Caro, G.M., Jim\'enez-Escobar, A., Mart\'in-Gago, J.A., et al. 2010, A\&A, 522, A108\\
Noble, J.A., Congiu, E., Dulieu, F., \& Fraser, H.J. 2012, MNRAS, 421, 768\\
Notesco, G. \& Bar-Nun, A. 2000, Icarus, 148, 456\\
\"Oberg, K.I. 2009, PhD Thesis, Universiteit Leiden\\
\"Oberg, K.I., Fayolle, E. C., Cuppen, H.C., van Dishoeck, E.F., Linnartz, H. 2009, A\&A, 505, 183\\
Pauls, T.A., Wilson, T.L., Bieging, J.H., \& Martin, R.N. 1983, A\&A, 124, 123\\
Prialnik, D., Sarid, G., Rosenberg, E.D., \& Merk, R. 2008, Space Sci. Rev., 138, 147\\
Sandford, S.A. \&  Allamandola, L.J. 1990, ApJ, 355, 357\\
Sandford, S.A. \&  Allamandola, L.J. 1993, ApJ, 417, 815\\
Sandford, S.A., Allamandola, L.J., Tielens, A.G.G.M., \& Valero, G.J. 1988, ApJ, 329, 498\\
Shu, F.H., Shang, H., \& Lee, T. 1996, Science, 271, 1545\\
Shu, F.H., Shang, H., Glassgold, A.E., \& Lee, T. 1997, Science, 277, 1475\\
Shu, F.H., Shang, H., Gounelle, M., Glassgold, A.E., \& Lee, T. 2001, ApJ, 548, 1029\\
Smith, R.S., Huang, C., Wong, E.K.L., \& Kay B.D. 1997, Phys. Rev. Lett., 79, 909\\
Snodgrass, C., Tubiana, C., Bramich, D.M., et al. 2013, A\&A, 557, A33\\
Thomas, P.C., Veverka, J., Belton, M.J.S., et al. 2007, Icarus, 187, 4\\
Viti, S. \& Williams, D.A. 1999, MNRAS, 310, 517\\
Viti, S., Collings, M.P., Dever, J.W., \& McCoustra, M.R.S. 2004, MNRAS, 354, 1141\\
Wakelam, V., Caselli, P., Ceccarelli, C., Herbs, E., \& Castets, A. 2004, A\&A, 422, 159\\
Wyckof, S. 1982, Comets, ed. L.L. Wilkening, (The University of Arizona Press, Tucson), 20\\
Yamada, H. \& Person, W.B. 1964, J. Chem. Phys., 41, 2478\\
Zubko, V., Dwek, E., \& Arendt, R.G. 2004, ApJSS, 152, 211\\

\end{document}